\documentclass[aps,prd,twocolumn,showpacs,superscriptaddress,groupedaddress, nofootinbib]{revtex4}  

\usepackage{graphicx}  
\usepackage{dcolumn}   
\usepackage{bm}        
\usepackage{amssymb} 
\usepackage{amsmath}
\usepackage[dvipsnames]{xcolor}
\usepackage{booktabs}
\usepackage{caption}
\usepackage{subcaption}
\usepackage{float}
\usepackage[normalem]{ulem}
\usepackage[hidelinks]{hyperref}

\definecolor{frenchlila}{rgb}{0.53, 0.38, 0.56}

\begin{document}

\title{Neural density estimation for Galactic Binaries in LISA data analysis.}
\author{Natalia Korsakova}
\email{korsakova@apc.in2p3.fr}
\affiliation{
 Astroparticule et Cosmologie, CNRS, Universit\'e Paris Cit\'e, F-75013 Paris, France
}%
\author{Stanislav Babak}
\affiliation{
 Astroparticule et Cosmologie, CNRS, Universit\'e Paris Cit\'e, F-75013 Paris, France
}%
\author{Michael L. Katz}
 \affiliation{NASA Marshall Space Flight Center, Huntsville, Alabama 35811, USA}
 \affiliation{Max-Planck-Institut f\"ur Gravitationsphysik, Albert-Einstein-Institut,  Am M\"uhlenberg 1, 14476 Potsdam-Golm, Germany}
\author{Nikolaos Karnesis}%
\affiliation{%
 Department of Physics, Aristotle University of Thessaloniki, Thessaloniki 54124, Greece
}%
\author{Sviatoslav Khukhlaev}
\affiliation{Department of Physics, Bar-Ilan University, Ramat-Gan, 52900, Israel}
\author{\\Jonathan R. Gair}
\affiliation{Max-Planck-Institut f\"ur Gravitationsphysik, Albert-Einstein-Institut,  Am M\"uhlenberg 1, 14476 Potsdam-Golm, Germany}

\begin{abstract}

The future space based gravitational wave detector LISA (Laser Interferometer Space Antenna) will observe millions of Galactic binaries constantly present in the data stream.  A small fraction of this population (of the order of several thousand) will be individually resolved.
One of the challenging tasks from the data analysis point of view will be to estimate the parameters of resolvable galactic binaries while disentangling them from each other and from other gravitational wave sources present in the data. This problem is quite often referred to as a \emph{global fit} in the field of LISA data analysis. 
A Bayesian framework is often used to infer the parameters of the sources and their number. The efficiency of the sampling techniques strongly depends on the proposals, especially in the multi-dimensional parameter space.
In this paper we demonstrate how we can use neural density estimators, and in particular Normalising flows, in order to build proposals which significantly improve the convergence of sampling.
We also demonstrate how these methods could help in building priors based on physical models and provide an alternative way to represent the catalogue of identified gravitational wave sources.
\end{abstract}

\maketitle

\section{Introduction}


Galactic ultra-compact white-dwarf binaries (GBs) are expected to be the most numerous gravitational wave (GW) sources for LISA (Laser Interferometer Space Antenna)~\cite{amaroseoane2017laser, Amaro_Seoane_2023}. 
We anticipate to observe millions of such systems~\cite{2001A&A...365..491N, Korol_2017}, 
 most of them will not be resolvable and their incoherent superposition will form a stochastic GW signal~\cite{Ruiter_2010}.  
 However, we should be able to individually resolve and characterize tens of thousands of such sources. 
The LISA data will contain tight GBs with periods ranging from a few minutes to about an hour. Those slowly inspiralling binaries will remain almost monochromatic over the several years that LISA will make observations. 

Such systems have been already detected via electromagnetic observations, and a fraction  of them will be directly detectable by LISA, thus being useful as \emph{verification} binary systems~\cite{Stroeer_2006, Kupfer_2018}. 
More of such binaries will be discovered in current GAIA~\cite{gaia} and ZTF surveys~\cite{Bellm_2019} or in future observations with LSST (Vera Rubin)~\cite{lsst}. Verification binaries can be used to monitor the performance of the instrument by validating parameters known from electromagnetic observations with GW observations.


The plus and cross polarisations of the time domain GW waveform in the Solar System Barycentre (SSB) are given by expressions~\cite{Babak:2021mhe}
\begin{align}
&h_+ (t) = A (1+\cos^2 \iota) \cos \Phi (t), \\
&h_\times (t) = 2 A \cos \iota \sin \Phi (t),
\end{align}
where $\iota$ is the inclination angle between the line of sight and the orbital angular momentum, $A$ is the amplitude of the GW signal and $\Phi$ is its phase.
The amplitude remains constant over the LISA observation time and is given by
\begin{equation}
A = 2 \frac{\mathcal{M}_c^{5/3}}{D_{\text{L}}} (\pi f)^{2/3}.
\end{equation}
Here $\mathcal{M}_c$ is the chirp mass, which is related to the masses of the binary components via ${\mathcal{M}_c = (m_1 m_2)^{\frac{3}{5}}/(m_1+m_2)^{\frac{1}{5}}}$, $f$ is the GW frequency and $D_L$ is the luminosity distance of the binary. 
Note that we use geometrical units throughout the paper, $G=c=1$. 

The phase of the GW can be expanded up to the first derivative in frequency, $\dot{f}_0$, using a Taylor expansion, assuming that binaries have circular orbits (i.e., we are ignoring the eccentricity of the orbit):
\begin{equation}
\Phi(t) = \phi_0 + 2\pi \left(f_0 t + \dot{f}_0 \frac{t^2}{2} \right),
\end{equation}
where $f_0$ is the initial frequency of the binary and $\phi_0$ is the initial phase of the waveform at the beginning of the observations.


The derivative of the GW frequency to the leading order, assuming that the evolution of the binary is only driven by GW emission, is a function of the chirp mass and the frequency itself
\begin{equation}
\dot{f} = \frac{96}{5} \pi^{8/3} \mathcal{M}_c^{5/3} f^{11/3}.
\end{equation}
Therefore, if measured, it allows us to estimate the chirp mass which can be combined with the measured amplitude to estimate the distance to the binary.

Other parameters characterising the GW signal include the sky position in the ecliptic SSB frame, which is specified by the ecliptic latitude $\beta$ and ecliptic longitude $\lambda$. 
In addition, the polarization angle $\psi$ describes the conversion of the waveform from the source frame to the SSB frame.
Overall the waveform is parameterised by the following set $\{ A, f_0, \dot{f}, \phi_0, \iota, \psi, \beta, \lambda \}$.

The motion of LISA around the Sun will cause a sky-dependent Doppler modulation of the GW phase. In addition, the directional LISA sensitivity (antenna pattern), while quite broad, will create an annual modulation of the GW amplitude. 
Usually, we choose to deal with the waveform in the frequency domain~\cite{Cornish_2007}, where the signal can be seen as a delta function $\delta_T (f-f_{GW}) \approx 
T \text{Sinc}[T(f-f_{GW})] $ with side bands arising from the Doppler modulation offset by $k/\rm{1\, yr}$ where $k$ is an integer and $T$ is the observation duration.

The most challenging problem for LISA data analysis is the simultaneous detection of overlapping GBs. The difficulty arises from the very high number of sources present in the data, and consequently, the a-priori unknown number of signals that can be classified as individually resolvable. In addition, the unresolved binaries will generate a stochastic foreground signal, which will depend both on the underlying population of binaries, and on our capabilities to extract their GW signatures from the data~\cite{amaroseoane2017laser, Littenberg:2023xpl}. The LISA global fit analysis, often based on Bayesian inference algorithms, is designed to solve exactly this dynamical high-dimensional problem~\cite{Littenberg:2023xpl,Karnesis:2023ras}. As expected, global fit analysis methods require a significant amount of effort to configure the Bayesian algorithms, for example constructing the proposal and prior distributions for thousands of GBs~\cite{Littenberg:2023xpl}.

In this work, we focus on methodology which provides an easy way to include the information available ahead of or during running the global fit analysis.
The first problem which we concentrate on in this paper is building a physical prior.
We will do it by fitting, with machine learning (ML) techniques, a distribution to a realisation of a theoretical model for the amplitude-sky parameters of the Galaxy.
The second part of our work focuses on improving 
the efficiency of Bayesian sampling techniques which rely on the proposal density (transition probability). Convergence of the Markov chain Monte-Carlo (MCMC) algorithm and the autocorrelation of the chains strongly depend on the proposal, especially in the multi-dimensional parameter space. We will show how the ML methodologies that we have developed can improve the efficiency of the conventional Bayesian approach.  

The paper is structured as follows. In the next Section~\autoref{sec:GB_da}, we formulate the main difficulty of LISA data analysis with the focus on GBs. We highlight particular points of a commonly-used Bayesian inference technique that can be enhanced with ML methods resulting in improved efficiency.
In~\autoref{Sec:NDE} we describe Normalising flows as the main tool in three practical applications with the results provided in~\autoref{Sec: Results}. We give a short summary with a road map for future research in~\autoref{Sec:discussion}.

\section{Galactic binaries data analysis problem}
\label{sec:GB_da}

The problem of the simultaneous analysis of the complete population of GBs has been investigated within the LISA community for more than a decade~\cite{Vallisneri:2008ye, Littenberg:2011zg, Littenberg:2020bxy, Littenberg:2023xpl,Strub:2023zxl}. 
The main difficulty lies in detecting multiple overlapping signals while simultaneously estimating the stochastic GW foreground which is expected to be the dominant noise component between $0.3$ and $3~\mathrm{mHz}$~\cite{amaroseoane2017laser, Littenberg:2023xpl, stochlisa}. 

The main technique for parameter estimation of a resolvable GB is a Bayesian approach
implemented within 
Markov Chain Monte Carlo (MCMC) methods~\cite{Cornish:2005qw,bda_gelman}. The methodology begins with Bayes' theorem
\begin{equation}
p(\theta| \mathcal{D}) = \frac{p(\mathcal{D}|\theta)\pi(\theta)}{p(\mathcal{D})},   
\end{equation}
where $p(\theta| \mathcal{D})$ is the posterior, $p( \mathcal{D}|\theta)$ is the likelihood, $p( \mathcal{D})$ is the evidence, and $\pi(\theta)$ is the prior of the parameters.
The Bayesian framework allows us to incorporate any knowledge that we have on parameters before the measurement into the prior distribution. 

The posterior probability for the given parameter set can be sampled by constructing a Markov chain by proposing new points in the parameter space, and deciding whether to accept or reject them based on an acceptance ratio that depends on the value of the posterior at the current and proposed points and is constructed to ensure that, asymptotically, the samples are drawn from the target distribution. In the widely-used Metropolis-Hastings method the acceptance probability is given by
\begin{equation}
    A(\theta', \theta) = \min\left( 
    1, \frac{p( \mathcal{D}|\theta') \pi(\theta') g(\theta|\theta')}{p( \mathcal{D}|\theta) \pi(\theta) g(\theta'|\theta)}
    \right),
\end{equation}
where $g(\theta'|\theta)$ is the proposal probability used
to make a jump to the new point $\theta'$ from $\theta$. In building the chains we rely on having a good proposal distribution. This is an essential part of the algorithm because it greatly affects the rate of convergence. 

Normally, the detection and parameter estimation of a single GB is a relatively easy task, but the problem becomes much more complicated when we have to deal with thousands of signals. The complexity arises from the high level of overlap between the signals, which makes their parameters correlated and requires a simultaneous joint fit for the unknown number of resolvable signals. For the case of LISA, and since the GW signatures of the GBs are very well localized in frequency, the analysis is usually performed in narrow frequency bands in parallel. This approach reduces the number of sources that have to be searched simultaneously, 
to the order of 10~\cite{Littenberg:2023xpl}. 
The unknown number of resolvable signals present in each frequency band elevates the problem to one of model selection, where the most appropriate model complexity, out of a set of models $\mathcal{M}_i$ (with $i$ being the number of sources in the data), needs to be determined simultaneously with the model parameters. Moreover, as already mentioned in the introduction, around the $\mathrm{mHz}$ band we expect a dominant stochastic component of the noise originating from the ensemble signal of the weak and unresolvable binaries. This increase in the uncertainty of the noise level adds another challenge to the analysis. So far, there have been several strategies proposed for analyzing such challenging data sets. We are going to incorporate our approach into two methods which we describe below.



\paragraph{Transdimensional Markov Chain Monte Carlo.}

The first approach focuses on transdimensional (or reversible jump (RJ)) samplers, which are essentially a generalization of the standard MCMC algorithms~\cite{green95, Littenberg:2023xpl, Karnesis:2023ras}. In RJMCMC, the dimensionality of the given model is also a parameter to be determined from the data. In practice, this means that we need to add an extra step to the standard MCMC algorithm, which proposes a model dimensionality reduction or increase (a parameter ``birth'' or ``death''). In the context of LISA data, this procedure is more straightforward, because the models are ``nested'', which means the more complicated model contains more GB signals compared to the simpler model. Thus, in the end we essentially need to propose to add or remove GB waveform signals during the sampling process.
 
However, the algorithm efficiency is highly dependent on the assumed prior, as well as the proposal distributions. The latter is crucial in achieving convergence within reasonable time scales. A convenient practice is to choose the proposal to be the same as the prior, because it simplifies the reversible jump acceptance ratio computation~\cite{Littenberg:2023xpl, Karnesis:2023ras}. Nevertheless, very often it can be proven to be inefficient, especially in cases like the one of LISA, which requires high-dimensional models and a broad prior which needs to cover the entire sky. In practice, the birth of a new source has very little chance of being accepted by drawing from the prior. To alleviate this problem, one can make more informative guesses about the model order and parameters based on a pilot run (focusing on detection or search), or even use techniques such as parallel tempering~\cite{Vousden2016}, which naturally come at an increased computational cost~\cite{Karnesis:2023ras}. Given the above, it is evident that as we build a model with a higher number of sources, it becomes critical to have an efficient proposal (which yields a high acceptance rate\footnote{As a reference, for a $d$-dimensional normal likelihood and a standard Metropolis-Hastings algorithm, an optimal rate of convergence is of the order or $\sim2.38/\sqrt{d}$~\cite{Gelman1996}, where the convergence rate is a time it take the chain to converge from an arbitrary point to a stationary distribution.}) for already added sources 
to overcome growth in dimensionality, to avoid adding the sources that should be already taken into account. 

\paragraph{Product space.} This method compares a constant (not dynamic as above) number of models~\cite{Hee, Bartolucci}. It introduces a hyperparameter $\kappa$ which enumerates models ($\mathcal{M}_i$ each containing ``$i$'' sources) and is used as one of the parameters in MCMC. This can be seen as a pine tree: each model corresponds to a branch with a fixed number of sources and $\kappa$ runs along the trunk jumping from branch to branch. The probability of a given model  (like in the transdimensional MCMC) is proportional to the length of the chain spent in the particular branch corresponding to that model.
The product space method can be seen as a static version of RJMCMC with a rule to jump between models. 
The robustness of this method strongly depends on the ability of the chain to explore each model, which becomes increasingly difficult with a large number of dimensions and a large number of models. Thus it is essential here to have a good proposal for each model.



\section{Density estimation with Normalising flows}
\label{Sec:NDE}

As we have argued above, in all MCMC-based methods good proposals are essential to efficiently build the sampling chain. More generally, we often want to build a continuous distribution based on available samples. 
There are multiple ways to approach this task. We can try to fit simplified models, like a Gaussian mixture model or Kernel Density Estimator~\cite{Falxa:2022yrm}.
However, such methods are rather sensitive to the choice of parameters and do not provide a generic enough density fit in dimensions higher than three. 

An alternative way to solve this problem is to use a form of the Neural Density Estimator (NDE), such as, for example, Invertible flows (or Normalising flows)\cite{Kobyzev_2021, JMLR:v22:19-1028}, which can provide a fit to an arbitrary distribution of samples and a corresponding measure of the probability.

The basic idea behind this approach is that we have a distribution that is simple to sample, such as, for example, a Normal distribution (hence the name), which we can transform to the target distribution. We represent the transformation by a sequence of invertible and differentiable mappings that we want to optimise to give the best possible representation of the target distribution. We usually call this simple distribution the {\it base} distribution. 
The map has to be invertible because the optimisation is done by transforming a sample from the distribution that we want to estimate 
back to the original base distribution and then computing the density of that inverse-transformed sample multiplied by the change in the volume of the transform.
This can be formalised as the change of variable equation
\begin{align}
p_X(\mathbf{x}) & =  q_Z(\mathbf{z}) 
| \det J_{f}(\mathbf{z}) |^{-1} \\
& = q_Z(f^{-1}(\mathbf{x})) 
| \det J_{f^{-1}}(\mathbf{x}) | \label{Eq:Main_NF}
\end{align}
where we define the transform $\mathbf{z} = f^{-1}(\mathbf{x})$. Variable $\mathbf{z}$ is drawn from a tractable distribution, $q_Z$, (which we call {\it base} distribution) and variable $\mathbf{x}$ is drawn from the distribution that we want to estimate, $p_X$. The Jacobian $J_{f^{-1}}$ determines the change in volume of the transform and is equal to the matrix of partial derivatives of the transform $f^{-1}$ relative to the variable $\mathbf{x}$. 

The idea behind a Normalising flow is to parameterise the transform $f$ in terms of a neural network. This allows for the possibility of capturing arbitrarily complicated distributions.

In this work, we use Normalising flows for density estimation. We start with 
the observed set of $N$ points from the distribution $\mathcal{D} = \{\mathbf{x}_i\}_{i=1}^N$. Then we can fit the density to a set of these points by maximising the log-likelihood of the data with respect to the parameters of the transform $\boldsymbol{\lambda}$: 
\begin{align}
   & \log p(\mathcal{D}|\boldsymbol{\lambda}) 
     = \sum_{i = 0}^N \log p_X(\mathbf{x}_i|\boldsymbol{\lambda}) \\
    & = \sum_{i = 0}^N \log q_Z(f^{-1}(\mathbf{x})|\boldsymbol{\lambda}) + 
    \log |\det J_{f^{-1}}(\mathbf{x}|\boldsymbol{\lambda})|,
\end{align}
where the parameters of the transform, $\boldsymbol{\lambda}$, in our case, are the weights of the neural network. After we optimise the network, we can invert the function $f$ and use it to produce samples from the distribution that we have fitted 
(as in the second line of \autoref{Eq:Main_NF}).

To define the transformation we have to fulfil some conditions. It was already mentioned that it has to be invertible and differentiable.
In addition, it is important that the Jacobian can be inverted in a reasonable amount of time ($<< \mathcal{O}(n^3)$), where $n$ is the dimension of the parameter space.
There has been a lot of progress in this direction in the field of ML where multiple approaches to construct the mapping have been proposed. The ones that we are going to use here 
are a combination of the NSF (Neural Spline Flows)~\cite{NEURIPS2019_7ac71d43} and RealNVP (Non-Volume Preserving flows)~\cite{dinh2017density}.

In the following subsections we describe several applications where we intend to use Normalising flows.

\subsection{Density fit for the Galaxy}
\label{subsec:prior_galaxy}

The density fit for the Galaxy, i.e., of the three parameters representing the position of a GB in space $\{ A, \beta, \lambda \}$,
can be used either as the prior or as a proposal. This will provide different ways to accommodate the fact that the spatial distribution of GBs is not uniform.
Models for the structure of our Galaxy, including the distribution of GBs, can be obtained from  
population synthesis calculations. 
These 
calculations evolve the progenitors of compact binaries based on the initial stellar population, star formation rate and Milky Way potential. There are several models available in the literature (e.g. ~\cite{2001A&A...365..491N, 2010A&A...521A..85Y, Lamberts_2019, Korol_2020, Breivik_2020}) which take into account different physical assumptions with various levels of complexity. Unfortunately, the astrophysical uncertainties are still quite large, 
however, the ultimate goal is to constrain these hypotheses/models with GW measurements by future space observatories (e.g as in~\cite{adams2012}).

Population synthesis models provide us with a catalogue of GBs with associated parameters: masses, orbital periods, sky positions and distances. Using those sources we simulate  
LISA data by adding extra parameters -- inclination, polarization and phase, drawn from uniform distributions. 

To set up an MCMC algorithm for parameter estimation we need a prior on all parameters. At the same time, we need to define proposals for the Markov Chain. In both cases, we can benefit from the knowledge on the spatial distribution of GBs in the Galaxy and construct the joint distribution for ($A$, $\beta$, $\lambda$) using a Normalising flow by fitting samples from a Galaxy model.


Different population synthesis codes will produce different models of the Galaxy. This should be kept in mind because if we choose a particular model we can end up with a prior which is too restrictive. Nevertheless, we have a relatively good knowledge on the shape of our Galaxy and samples from any given model will provide a good description of spiral galaxies with structures close to the Milky Way. 
In this work, we demonstrate how to build a fit to the model of the Galaxy used in the simulated LISA data ``Sangria''~\cite{le_jeune_maude_2022_7132178}. This catalogue of sources was drawn from the  Galaxy model based on the population synthesis described in~\cite{Korol_2020}.
We emphasise, however, that the procedure described in this paper could be used for any other Galactic model. 

In addition, we also try to build a ``generic'' model of the Galaxy, by making the distribution of binaries rather broad. This will allow us to relax the dependency on the particular model we have chosen for the fit and make it smooth enough to accommodate a wider range of possible galactic shapes.

For the generic Galaxy, we use the same distribution from the ``Sangria'' dataset but smoothen it by taking the logarithm of the probabilities. In practice, this is done in the following way: 
\begin{enumerate}
    \item Build a binned 3-dimensional histogram of the ``Sangria'' Galactic model. 
    \item Modify the number of sources in each bin: \begin{itemize}
        \item if there is 0 or 1 source in the bin, it remains unchanged,
        \item if there are $N \geq 2$ sources in the bin, we replace it with $\lfloor\log N\rfloor$ sources.
    \end{itemize}
    \item Construct new Galactic samples by drawing from the newly built 3-d binned histogram. 
\end{enumerate}
The new ``broad'' Galaxy could be used in the analysis and the posterior samples then re-weighted for a particular Galactic model within a hierarchical Bayesian framework.

\subsection{Density fit for posterior distributions}

The second important application of the density fit is building proposal probability distributions. As discussed in~\autoref{sec:GB_da} a good proposal is essential for an efficient MCMC because it will result in a high acceptance rate and low autocorrelation of the chain. 
There are two cases where we want to convert previously obtained posterior distributions into proposals. 


The first case is related to the time iterative analysis of LISA data. GBs are present in the data all the time therefore we are updating the data volume as more observations become available with a certain cadence. Preliminary assessments indicate that using a 2-4 month cadence of the data update is optimal for GBs~\cite{stas_globalfit}. 
As we grow the data volume we want to use the results obtained previously, in one of two ways: (i) use the old posterior as a prior for the new data; (ii) use the old posterior to form a proposal for the extended data. In this work, we explore the second possibility.  Note that the time iterative analysis also works as a natural annealing scheme, we first detect the strongest sources and uncover more sources as we accumulate more data (the signal-to-noise (SNR) for these sources grows roughly as the square-root of the observing time). In particular, this also implies that the dimensionality (the number of sources) could vary after each data update. We suggest using NDE to fit the posterior distribution that can be used as a proposal. More precisely we build the proposal based on the fit to the posterior distribution performed with the Normalising flow. This fit allows us to sample from the estimated distribution and for each point in the distribution it also returns the value of the log probability that can be directly used as the proposal probability.
One caveat here is that we tend to find the nearby sources in the early data, so if we build a proposal for the amplitude (and to a lesser extent sky localisation) for the first 2 months of data it won't have good coverage of the sources we could be seeing with 2 years of data. 

Builing such proposals can be also useful in the case of parallel tempering. Parallel tempering is a technique that is based on running multiple MCMC chains in parallel that are sampling different tempered likelihoods. This means that for high temperatures the log-likelihood is being heated and smoothed, allowing for efficient sampling of the complete parameter space. Information about the states of the chains at each temperature is then propagated towards the cold chain. If parallel tempering is used in the analysis we can build additional proposals based on the samples from one of the hot chains, which gives a somewhat wider distribution accounting for possible inaccuracy but preserving all correlations between parameters. The proposals based on the cold and hot chains could be used in alternation or as a part of the larger set of proposals.

The second case is related to the unknown number of sources and building an $N$-source model. It is convenient first to identify possible sources in a given frequency band solving the problem of ``detection''.  We will not describe in detail how we detect GBs here, but just outline the main method. We first identify the loudest signals with a preliminary estimation of their parameters, then we subtract these sources and search for the weak(er) signals. The identified GB candidates could be used as a proposal for the transdimensional MCMC or in product space. Note that we carefully call them ``candidate'' GBs as they could be spurious due to a strong correlation between the overlapped sources and/or due to low SNR. We also do not discuss any SNR threshold for the detection, as we let Bayesian model selection take care of it, however, eventually, we will still need to decide on the threshold for the Bayes factor. In this paper we demonstrate the procedure of using density fits to build the proposal probability for GBs, nevertheless, the same approach could be used for other GW sources.

\subsection{Alternative representation of GW catalogues}
\label{sec:alternative_catalog}

The results of parameter estimation, as produced by the LISA data analysis pipelines, 
have to be shared with the community of Astronomers. In the case of Galactic binaries, this will be $\sim \mathcal{O}(10^4)$ sources. It is anticipated that GW catalogues will contain point estimates for the parameters of each source and posterior probability estimates of parameter uncertainties. The latter will be challenging due to the large number of overlapping signals in certain frequency bands, in this case marginalised posteriors for each individual source might be missing important information.
Therefore we will have to store joint or marginalized posteriors. Working within a Bayesian framework, we will have posteriors represented by a set of samples.
Thus we will have to decide on how many samples are enough to encapsulate the correlation between parameters. 

We propose an alternative way to store the posteriors: we can transform posterior distributions using NDE and store the object (with metadata) that could be used to draw as many samples as an end-user wants. Those can be seen as smooth (semi)analytic fits to the observed posteriors.

\section{Results}
\label{Sec: Results}

In this section, we will describe the practical implementation of the NDE in a form of the Normalising flow and then show the results of applying it to fit the distributions described in the previous Section.

\subsection{Details on the network}

We base our calculations on the \verb+lisaflow+ package that we have developed for the parameter estimation of LISA data using Normalising flows (publication in preparation). The architecture of the flow consists of two types of coupling flows with two different designs for the coupling layers: Neural Spline Flows (NSF)~\cite{NEURIPS2019_7ac71d43} and Real Non Volume Preserving flows (Real NVP)~\cite{dinh2017density}. 
The implementation of the flow part of the package heavily relies on the one described in~\cite{nflows}. 

\begin{figure}[!h]
\includegraphics[width=\columnwidth]{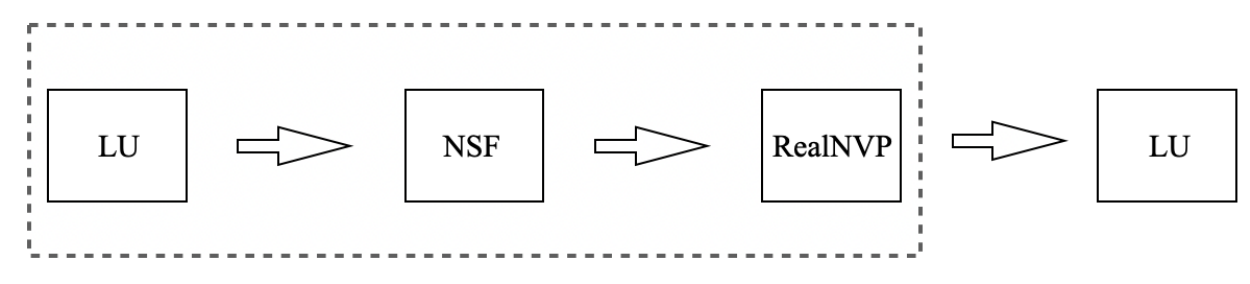}
    \caption{Schematic representation of the network architecture. Each block represents a component of the flow. The dotted line describes the bloc that is repeated multiple times. LU is a linear transform implemented using lower-upper decomposition; NSF is a neural spline flow; and RealNVP is a real non volume preserving flow.}
    \label{Fig:NF}
\end{figure}

\autoref{Fig:NF} gives a schematic representation of the flow architecture. 
It is a repetitive combination of blocks which are used to model bijective transformations. Each block contains a combination of the linear transform (using lower-upper decomposition -- LU), piecewise rational quadratic coupling transform (NSF) and affine transform (RealNVP). Within each block, the functions performing the transform are modeled using neural networks. We use a residual network (ResNet)~\cite{he2016deep} for the network architecture.

It is important to optimise the size of the network and the time it takes for the network to train when we fit the posterior distributions. NDE indeed provides efficient proposals, but we do not want this to be counterbalanced by long training times. In addition, the training time is important because we might need to build fits for $\sim 10000$ signals. 

In our implementation we have taken the number of repetitive flow blocks to be equal to 4. 
For both flow blocks we use a ResNet with 256 hidden features, a ReLU activation function, a dropout probability of 0.2 and batch normalisation. The number of ResNet layers in NSF is 4 and in RealNVP is 2.


In the following subsections, we give the results of NDE for each application described above, maintaining the same order. 

\subsection{Fit to the Galaxy}

\begin{figure*}[!t]
\begin{subfigure}[t]{0.45\textwidth}
\centering 
\includegraphics[width=\textwidth]{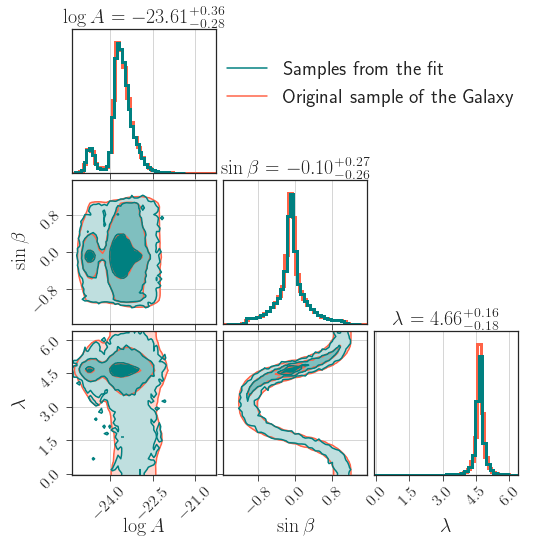} 
\caption{Fit to the ``Sangria'' population of GBs. We have used three-dimensional parameter space: amplitude and ecliptic sky coordinates. \label{fig_galaxy_original}}
\end{subfigure}
\begin{subfigure}[t]{0.45\textwidth}
\includegraphics[width=\textwidth]{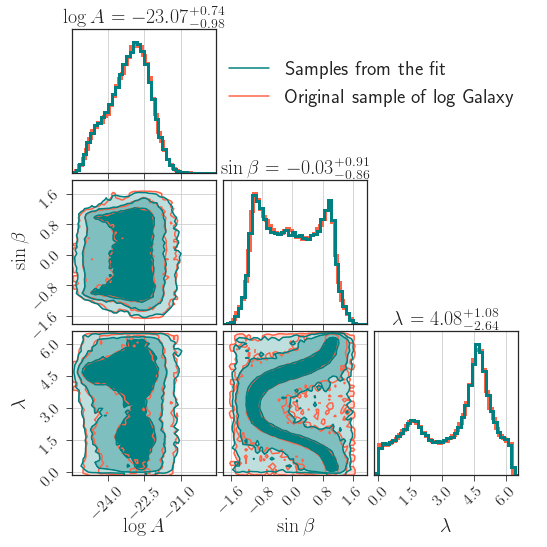}
\caption{Fit to the samples of the ``generic'' Galaxy
obtained by broadening the ``Sangria'' GB distribution, for details see description in \autoref{subsec:prior_galaxy}.\label{fig_galaxy_log}}
\end{subfigure}
\caption{Fit to the Galaxy. Orange -- original sample, cyan -- inverse transform of sample from the normal distribution. Contours represent the lines of the equal probabilities for the values of 0.68 0.95 and 0.997.}
\label{Fig:res_galaxy}
\centering
\end{figure*}

As stated previously we have used a ``Sangria'' catalogue of GBs to train the neural network. 
The catalogue contains more than 30 million GBs. We have combined GB parameters from the sky localisation and observed GW amplitude to work with a three-dimensional parameter space of sky-amplitude $A, \beta, \lambda$. The results are presented in \autoref{Fig:res_galaxy}. 

The left panel compares the distribution of parameters of GBs from the catalogue (orange) and the distribution obtained from sampling the trained Normalising Flow (dark cyan).
It is evident that the two posterior probability density functions are almost identical.
The right panel compares the distribution of the ``broad'' Galaxy, where we have used similar colours. Again we observe an exceptional  performance of the Normalising flow. The training for the Galaxy has to be done once before the analysis of the data, so we can do an extended training without being too concerned about the time it takes. 
Depending on the number of outliers that we consider acceptable the training can take from 10 minutes up to several hours.

\subsection{Fit to the posterior}

\begin{table*}[t]
\begin{ruledtabular}
\begin{tabular}{lllllllll}
Source & $A$ & $f$ & $\dot{f}$ & $\beta$ & $\lambda$ & $\iota$ & $\phi$ & $\psi$ \\
\colrule
AMCVn  & $2.829116\cdot 10^{-22}$  & $0.001944$  & $6.061897\cdot 10^{-18}$ &
$0.653496$ & $2.973723$ & $0.750492$ & $5.141845$ &  $3.567122$  \\ 
ESCet & $1.068882\cdot10^{-22}$ & $0.003225$ & $-8.319025\cdot 10^{-18}$ &
$-0.354893$ & $0.429492$  & 
$1.047198$ & $5.624013$ &  $1.155596$  \\
V803Cen & $1.599302\cdot 10^{-22}$ & $0.001253$ & $1.064206\cdot 10^{-18}$ & 
$-0.529128$ & $3.772765$ &
$0.235619$ & $2.941195$ &  $4.698031$  \\
ZTFJ1539 & $9.920754\cdot 10^{-23}$ & $0.004822$ & $2.537825\cdot 10^{-16}$ &
$1.154738$ & $3.578408$ & 
$1.468695$ & $4.830600$ & $5.088202$  \\
\end{tabular}
\end{ruledtabular}
\caption{Parameters of the verification binaries that were selected for testing the performance of the algorithm.}
\label{tab:VBs}
\end{table*}

\begin{figure*}[hbtp]
\begin{subfigure}{0.45\textwidth}
\centering 
\includegraphics[width=\linewidth]{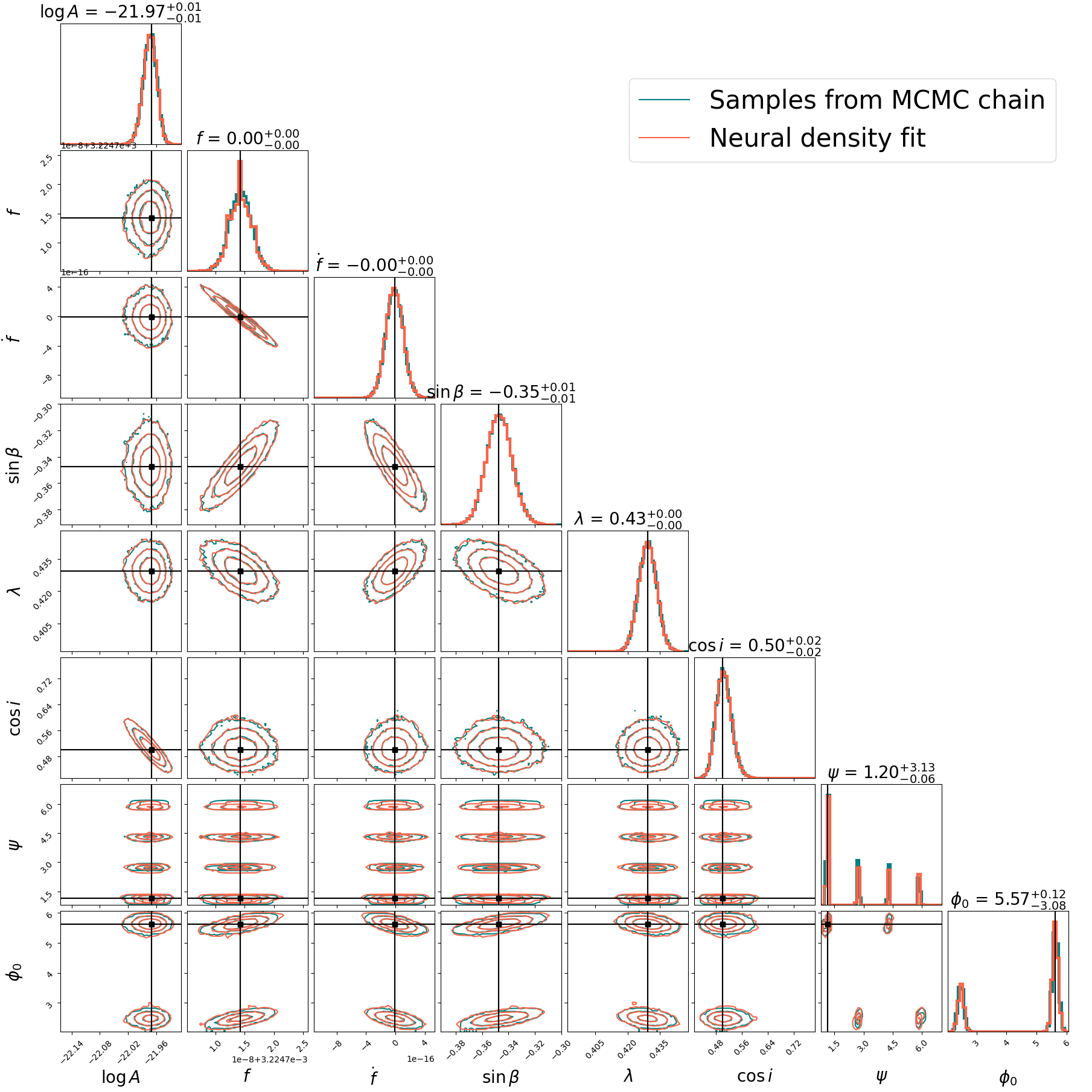} 
\label{fig_ESCet}
\end{subfigure}
\hfill
\begin{subfigure}{0.45\textwidth}
\includegraphics[width=\linewidth]{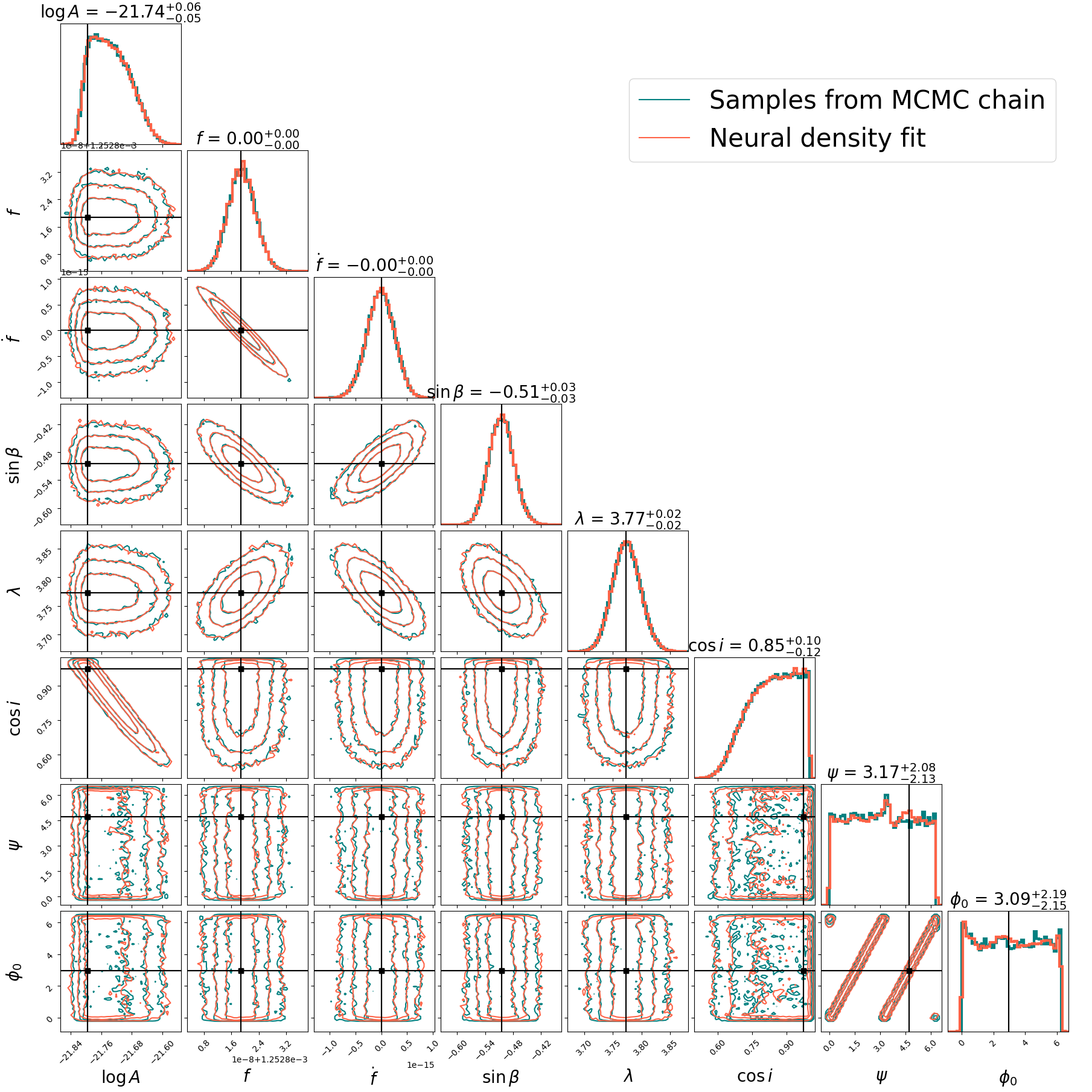}
\label{fig_V803Cen}
\end{subfigure}
\bigskip
\begin{subfigure}{0.45\textwidth}
\includegraphics[width=\linewidth]{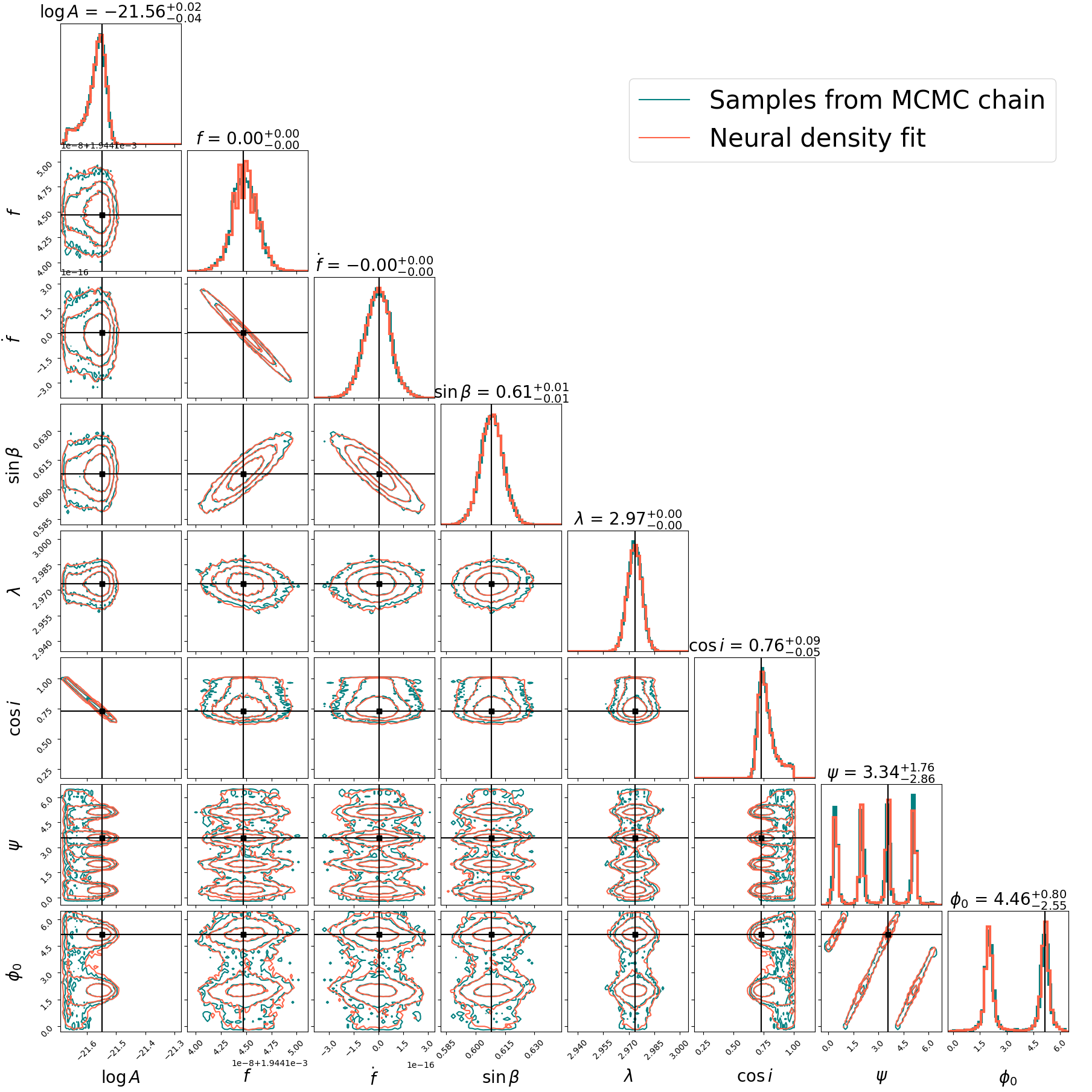} 
\label{fig_AMCVn}
\end{subfigure}
\hfill
\begin{subfigure}{0.45\textwidth}
\includegraphics[width=\linewidth]{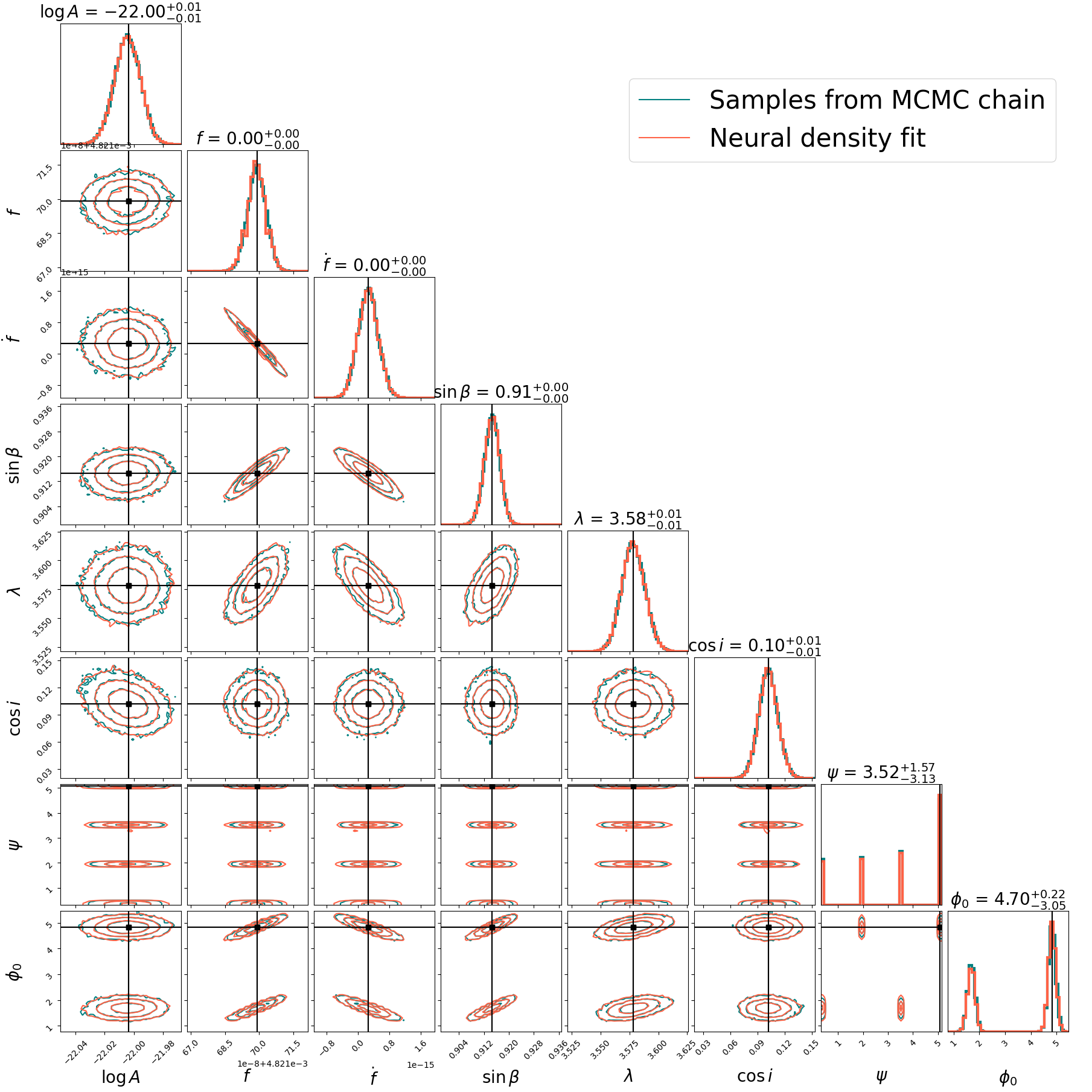}
\label{fig_ZTFJ1539}
\end{subfigure}
\caption{Comparison of NDA and MCMC posteriors for ESCet (top left), V803Cen (top right), AMCVn (bottom left) and ZTFJ1539 (bottom right). The contours correspond to the levels of the equal probability with the chosen values of 0.68 0.95 and 0.997. The cyan colour corresponds to the MCMC samples from the original distribution 
while the orange colour corresponds to samples from the neural density fit performed with the Normalising flow. In all cases, the two sets of contours show excellent agreement.}
\label{Fig:posteriors}
\centering
\end{figure*}

\begin{figure*}[hbtp]
\begin{subfigure}{0.45\textwidth}
\centering 
\includegraphics[width=\linewidth]{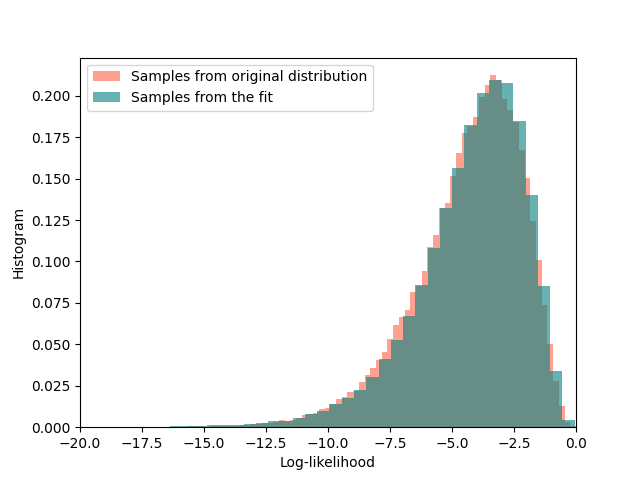} 
\caption{Loglikelihood distribution for ESCet after 1000 iterations.}
\label{fig_loglike_ESCet}
\end{subfigure}
\hfill
\begin{subfigure}{0.45\textwidth}
\includegraphics[width=\linewidth]{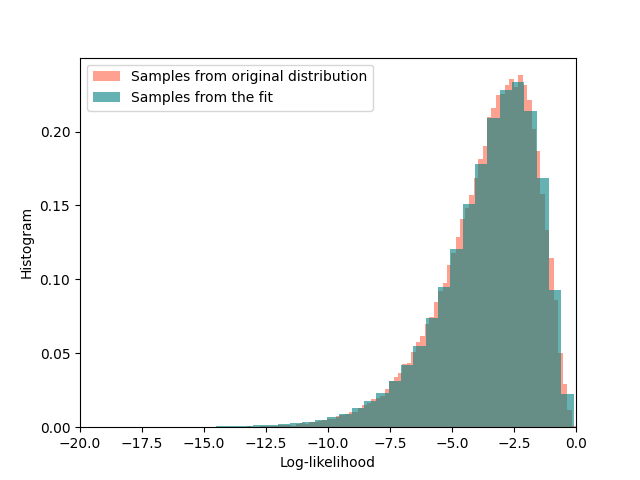}
\caption{Loglikelihood distribution for V803Cen after 1000 iterations.}
\label{fig_loglike_V803Cen}
\end{subfigure}
\bigskip
\begin{subfigure}{0.45\textwidth}
\includegraphics[width=\linewidth]{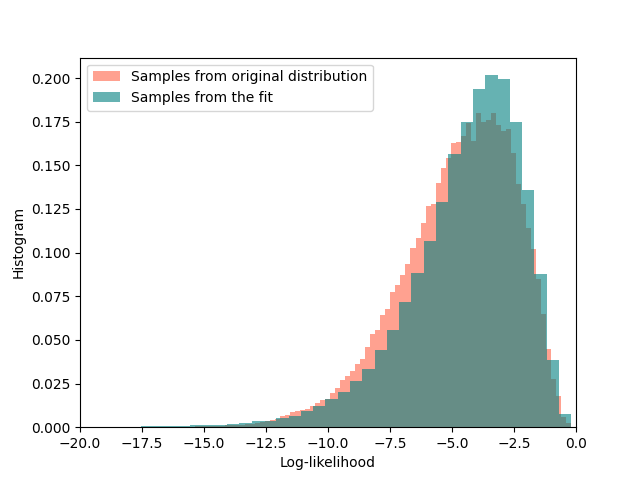} 
\caption{Loglikelihood distribution for AMCVn after 1000 iterations.}
\label{fig_loglike_AMCVn}
\end{subfigure}
\hfill
\begin{subfigure}{0.45\textwidth}
\includegraphics[width=\linewidth]{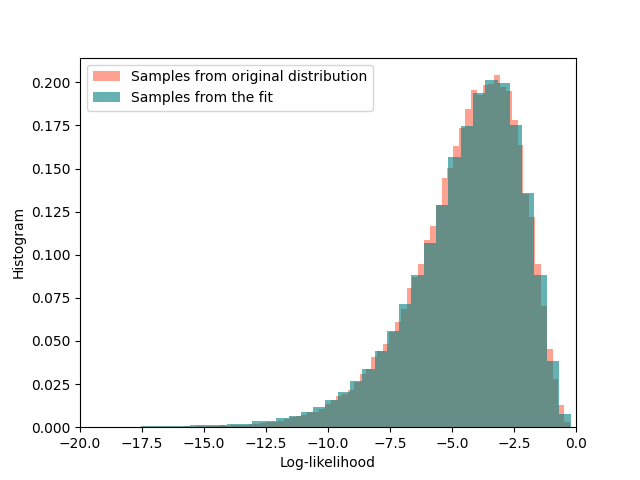}
\caption{Loglikelihood distribution for ZTFJ1539 after 1000 iterations.}
\label{fig_loglike_ZTFJ1539}
\end{subfigure}
\caption{Log-likelihood distribution for MCMC samples and the samples drawn from the NDE fit to the posterior, after 1000 iterations. The sub-panels show results for ESCet (top left), V803Cen (top right), AMCVn (bottom left) and ZTFJ1539 (bottom right).}
\label{Fig:loglik}
\centering
\end{figure*}

We have used four verification galactic binaries (VGBs), which are listed along with their parameters in~\autoref{tab:VBs}. All VGBs listed in this table are detectable and were chosen to cover the frequency band between 1 and 5 mHz. These sources were chosen to include both detached and interacting binaries. Interacting binaries experience mass transfer which is reflected in the parameter $\dot{f}$ that can take negative values. For each binary we have produced posterior samples using an MCMC implementation described in \cite{Falxa:2022yrm}. The posterior samples were used to train the NDE. First, we visually compare the distributions from MCMC and NDE for each source, with the results shown in \autoref{Fig:posteriors}. The two distributions are shown in cyan (MCMC) and orange (NDE) colours, but the agreement is so good that they are almost completely overlaid.        

\begin{figure}[!h]
\includegraphics[width=\columnwidth]{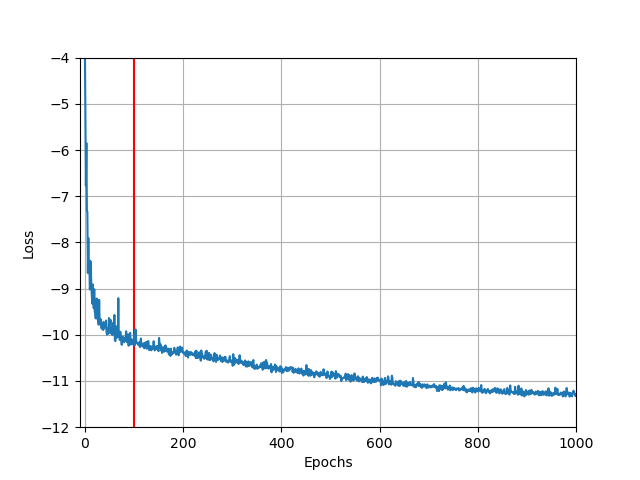}
\caption{Evolution of the loss function for the training of the fit to the posterior of ZTFJ1539. The X-axis represents the number of epochs (one epoch is one full iteration over all samples of the posterior) and the y-axis is the value of the loss. Red line represents $\text{epoch} = 100$.\label{Fig:loss}}
\end{figure}

The corner plot gives the two-dimensional projection of the posterior and does not fully reflect the multi-dimensional distribution. Therefore, we also assess the performance of the fit by looking at the distribution of likelihood values. For each source, we draw samples from the NDE fit and compute the likelihood. Note that we have used the ``noiseless'' likelihood: 
\begin{equation}
\log \mathcal{L} \propto -\frac1{2} 
(s-h(\theta) | s - h(\theta)),
\end{equation}
where $s$ is the signal, $h(\theta)$ is the template, and the inner product $(.|.)$ can be computed in either the time or frequency domain and is weighted by the noise spectral density evaluated at the frequency of the GB signal.


In \autoref{Fig:loglik}, we compare the distribution of the log-likelihood from the NDE samples with the distribution obtained from the MCMC chain. In all of the cases we observe excellent agreement between the two distributions, which implies that if we would use the NDE as a proposal in an MCMC run, we would expect high acceptance rate of this proposal, as it closely resembles the target distribution.
We note here that the acceptance rate can be also influenced by other factors which should be accessed separately. 

It is important to highlight that the performance of the NDE fit depends on the number of iterations used in training the network. Quoted results were obtained with 1000 iterations and the training process takes around 10 seconds for each epoch when using an NVIDIA V100 GPU. The total training time was approximately $2.7$ hours for the 1000 epochs used here. We have chosen this number of epochs because the evolution of the loss function shown in~\autoref{Fig:loss} shows that the loss is approaching convergence at this point.

\begin{figure}[h!]
    \centering
    \begin{subfigure}[t]{\columnwidth}
        \centering
        \includegraphics[width=\columnwidth]{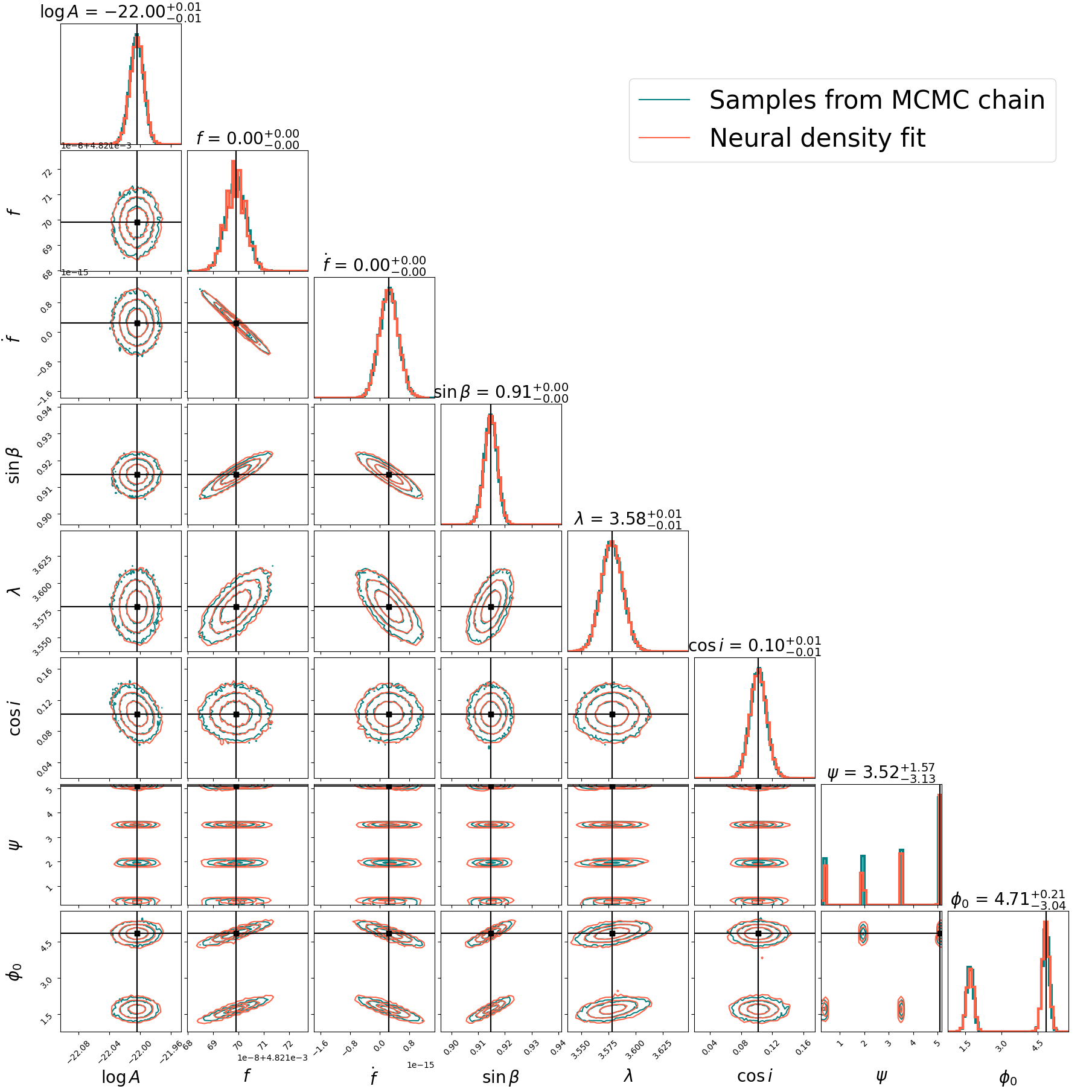}
        \caption{The corner plot shows very good agreement between the MCMC and NDE distributions for ZTFJ1539, even with truncated training. The contours correspond to levels of equal probability with the chosen values of 0.68, 0.95 and 0.997. The cyan contours correspond to samples obtained with MCMC, while the orange colour shows results from the neural density fit performed with the Normalising flow.}
    \end{subfigure}%
    \newline
    \begin{subfigure}[t]{\columnwidth}
        \centering
        \includegraphics[width=\columnwidth]{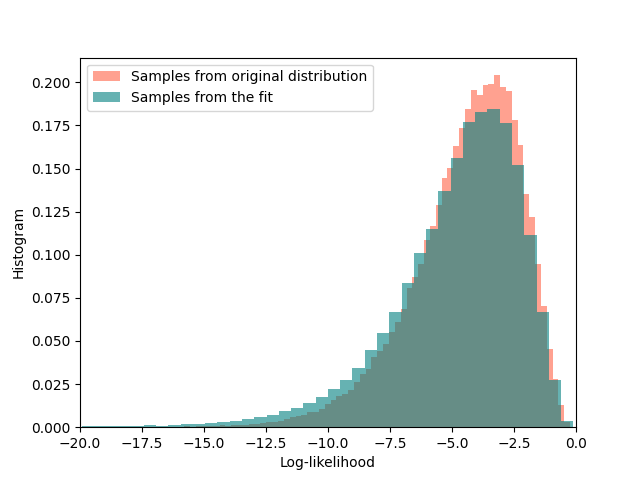}
        \caption{Loglikelihood distribution for ZTFJ1539 after 100 iterations.}
    \end{subfigure}
    \caption{Comparison of the likelihood distribution evaluated on the samples drawn from the NDE fit (cyan)  after 100 epochs of training and from the samples obtained by MCMC (orange). \label{Fig:epoch_100}}
\end{figure}

The quoted time which was used for training will be feasible in the case if we get the intermediate posteriors that we need to fit with some large cadence and we do not need update them very often. In the situation when we need to fit the posteriors on the flight while sampling the quoted time will be too long.
In this case we can trade accuracy in reproducing the full probability density for efficiency in training. To test this, we truncate the training after 100 epochs, marked in~\autoref{Fig:loss} as a red vertical line.
 Results obtained from using the resulting network, which are presented in~\autoref{Fig:epoch_100} for ZTFJ1539, show that at this stage the fit has already reached reasonably good convergence, demonstrated both by the corner plot and by the distribution of the likelihoods. We have gained a factor of 10 in efficiency (training for 100 epochs takes about 17 minutes) with only a slight drop in the accuracy of the fit.
Since we envisage using the NDE as a proposal within a sampling algorithm, it does not have to match perfectly,
so we have some freedom to choose an optimal operational point based on the efficiency-accuracy criteria.

 

\subsection{Sampling from the network}
 
In~\autoref{sec:alternative_catalog} we stated that the results of parameter estimation for astrophysical sources in the LISA data will have to be provided to the scientific community. We suggested that one of the possibilities is to use a fit to the posterior. Instead of sharing samples from the posterior distributions, we can provide the weights of a network that has been trained to fit the data along with code to sample from this network. This will allow the user to create an arbitrary number of samples and will be a convenient way to share the information.

For storing the catalogues in the form of NDE fits, we need to take into account the size of the network image.  The speed of the convergence depends on the size of the network: a bigger network (with a large number of weights) will converge faster. However, the size of the network is restricted by the memory of the GPU and by the size of the image we ultimately want to store. 

Storing too large images might be quite impractical, the size of the model that we have trained is about 10MB per source. The size of the image also depends on the particular form of NDE that is used, and could be further optimised. 

However, the output is really a joint posterior across all sources, and so will potentially include additional information characterising correlations that means it's final size will not just be linear in the number of sources.



\section{Discussion and future work}
\label{Sec:discussion}
In this work, we have shown how neural density estimation using Normalising flows can be used to represent probability distributions from which only samples are available. We have also discussed several applications of this approach to help improve sampling within a Bayesian analysis framework. The first was to represent a prior or a proposal constructed from a theoretical model of the population of GW sources. The second was to build a proposal distribution that can significantly improve the efficiency of MCMC algorithms. The third application 
was as a way to distribute the final results of parameter estimation of LISA sources to the scientific community.

These applications are being incorporated into the LISA global fit pipelines where we are simultaneously detecting/characterising several populations of GW sources and estimating the noise (both instrumental and the astrophysical foreground). In follow-up work, we will describe in detail the implementation and the gain in efficiency which we have achieved by using the NDE approach suggested in this paper.




\section*{Acknowledgements}
We would like to thank CC-IN2P3 cluster for providing computational resources.
S.B. and N.Korsakova acknowledge support from the
CNES for the exploration of LISA science. The participation of S.B. in this project has received funding from the European Union’s Horizon 2020 research and innovation program under the Marie Skłodowska-Curie grant agreement No. 101007855. N.Karnesis acknowledges the funding from the European Union’s Horizon 2020
research and innovation programme under the Marie Skłodowska-Curie grant agreement No 101065596.
S.B. acknowledges funding from the French National Research Agency (grant ANR-21-CE31-0026, project MBH\_waves).

\clearpage

\bibliography{bib_density}

\end{document}